\documentclass[aps, twocolumn, nofootinbib, showpacs, prl]{revtex4}
\usepackage{graphicx}
\usepackage{amsfonts}
\usepackage{amssymb}
\usepackage{amsbsy}
\usepackage{amsmath}
\usepackage{mathtools}
\usepackage{latexsym}
\usepackage{natbib}
\usepackage{bm}
\usepackage{ulem}
\usepackage[utf8]{inputenc}
\usepackage{newunicodechar}
\newunicodechar{∼}{$\sim$}
\usepackage{color}
\usepackage{float}
\usepackage{hyperref}
\usepackage{multirow}
\usepackage{array,makecell}
\begin{document}
\newcommand{\setone}{\textit{Pantheon Plus \ensuremath{+} CMB \ensuremath{+} DESI DR2}}
\newcommand{\settwo}{\textit{DES Y5 + CMB + DESI DR2}}
\newcommand{\setthree}{\textit{ CMB + DESI DR2}}
\title{Is Phantom Barrier Crossing Inevitable? A Cosmographic Analysis}

\author{Nandan Roy}
\email{nandan.roy@mahidol.edu}

\affiliation{Centre for Theoretical Physics and Natural Philosophy, Nakhonsawan Studiorum for Advanced Studies, Mahidol University, Thailand}

\author{Soumya Chakrabarti}
\email{soumya.chakrabarti@vit.ac.in}

\affiliation{School of Advanced Sciences, Vellore Institute of Technology, Vellore, Tiruvalam Rd, Katpadi, Tamil Nadu 632014, India}


\date{\today}

\begin{abstract}
Recent findings from the Dark Energy Spectroscopic Instrument (DESI), analyzed together with supernova observations and CMB measurements, provide statistically significant indications (at the 2-5$\sigma$ level) of a time-varying dark energy component alongwith a possible phantom-to-quintessence transition in the recent past. In this letter, we investigate the evolution of dark energy using a model-independent cosmographic approach and explore the possibility of phantom barrier crossing. By mapping the differential equation defining jerk parameter into an anharmonic oscillator, we derive an analytical expression for the dark energy equation of state (EoS), which, remarkably, depends on a single parameter. Using DESI-DR2 BAO data, supernova data, and a compressed Planck likelihood, we constrain the cosmological parameters and find deviations from a cosmological constant at late times. Unlike the CPL parametrization, our results show no phantom barrier crossing, highlighting the power of kinematic reconstructions in probing the nature of dark energy. Furthermore, using a dynamical system approach, we demonstrate that $w_{DE}=-1$ acts as a bifurcation point with degenerate stable fixed points and therefore prevents solutions from crossing this barrier from either side.
\end{abstract}

\maketitle

The fundamental nature of dark energy (DE) continues to remain unresolved \cite{PhysRevLett.86.6, PhysRevD.67.103508} even after decades since its postulation, following the confirmation of late-time cosmic acceleration by supernova redshift-luminosity surveys \cite{Riess_2001, Rubin_2016}. While the cosmological constant $\Lambda$ remains a popular candidate to fill in this role, the observed and theoretically predicted values are in substantial disagreement. On the other hand, a traditional approach like quintessence, often mimic the $\Lambda$CDM model in the present era but are challenged by contradictions with the equivalence principle \cite{PADMANABHAN2003235, Eisenstein_2005}. More contemporary models of dynamical DE allow for smooth transitions between different cosmological epochs, utilizing time-dependent equations of state (EoS) that also accommodate structure formation. Recent DESI Data (DR1 and DR2) \cite{DESI:2024mwx, DESI:2025zgx} favors a dynamical dark energy over $\Lambda$CDM and at the same time suggests that dark energy may have transitioned from a phantom into a quintessence phase \cite{Linder:2024rdj}. However, concerns remain regarding the potential limitations of the CPL parametrization, which has been adopted in the DESI analyses and this issue deserves further attention. Its' reported limitations include the lack of a strong physical basis, sensitivity to priors, and model-dependent biases that can artificially favor dynamical dark energy or disfavor $\Lambda$CDM at high statistical significance \cite{Carloni:2024zpl, cortes2024interpreting, Nesseris:2025lke, Sakr:2025daj, Notari2024Consistent,Dinda:2025iaq,Colgain:2025nzf}. \\

In this letter, we attempt to see through the evolution and the potential phantom barrier crossing of \textit{dark energy} using a model-independent approach based on kinematic parameters rooted in a cosmographic framework. The central element of this approach is the jerk parameter for which there are recent parametrizations \cite{luongo, Rapetti_2007, Zhai_2013} explored in literature, underscoring its importance in cosmological reconstructions \cite{Alam_2003, Alam_2004, sahni1, PhysRevD.60.081301}. The parameter naturally produces a non-trivial third-order differential equation for $a$, unless the jerk equals $-1$ (which produces $\Lambda$CDM). For a spatially flat, homogeneous and isotropic universe we find a way to reframe the differential equation defining the jerk parameter into an anharmonic oscillator equation. With a Painleve symmetry analysis, any such oscillator equation can be morphed into a total derivative, leading to a direct integration \cite{Euler1997, PhysRevD.95.024015, Chakrabarti2023}. In the process, we derive an analytical expression for the dark energy equation EoS, which remarkably depends on a single parameter. Using \textit{DESI-DR2} together with different supernova data compilations, and a compressed Planck likelihood analysis, we constrain the cosmological parameters and find deviations from a cosmological constant at late times alongwith a statistically significant preference for the dynamical dark energy. However, we do not find any phantom barrier crossing. To further investigate, we reformulate the kinematical equation for the jerk parameter into a dynamical system, for which the phantom barrier crossing acts as a bifurcation point. It essentially allows us to argue that, from a general dynamical system perspective, the so-called phantom barrier crossing isn't an unbending fate.  \\

\textit{Mathematical Formulation :}
In cosmography, the scale factor $a(t)$ is expanded using a Taylor series, from which key kinematic quantities are derived. The Hubble parameter $H = \frac{\dot{a}}{a}$ comes from the first derivative, the deceleration parameter $q = -\frac{\ddot{a}/a}{\dot{a}^2/a^2}$ from the second and the jerk parameter $j(t) = -\frac{1}{a H^3}\left(\frac{d^3 a}{d t^3}\right)$ from the third derivative. Redshift $z$ is defined as $z = \frac{a_0}{a} - 1$, with $a_0 = 1$. The jerk parameter can be expressed as a function of redshift

\begin{equation}\label{jerk1}
j(z) = -1 + (1+z) \frac{\left(h^2\right)'}{h^2} - \frac{1}{2} (1+z)^2 \frac{\left(h^2\right)''}{h^2},
\end{equation}
where $h(z) = \frac{H(z)}{H_0}$ and primes denote derivatives with respect to $z$. First, we rewrite this as
\begin{equation} \label{jerk2}
\phi'' - \frac{2\phi'}{(1+z)} + \frac{2(1+j)\phi}{(1+z)^2} = 0,
\end{equation}
where $\phi(z) = \frac{H(z)^2}{H_0^2}$. This equation resembles a classical anharmonic oscillator and enforcing a rescaling of $2(1+j) = \phi^{(n-1)}J(z)$, we obtain

\begin{equation}\label{integrable}
\phi'' - \frac{2\phi'}{(1+z)} + \frac{\phi^n J(z)}{(1+z)^2} = 0.
\end{equation}
This suggests that the evolution of the jerk parameter (and potentially others) is governed by the Hubble function, albeit involving the correction term $J(z)$. We apply Painleve symmetry analysis on this equation and transform it into an integrable form \cite{Euler1997, PhysRevD.95.024015, Chakrabarti2023}. The general form of an anharmonic oscillator equation is $\ddot{\phi} + f_1(t) \dot{\phi} + f_2(t) \phi + f_3(t) \phi^n = 0$, where $f_1$, $f_2$ and $f_3$ are functions of time, and $n$ is a constant. For the equation to be integrable, we require the condition \(n \notin \{-3, -1, 0, 1\}\) and a constraint on the coefficients
\begin{eqnarray}\label{criterion1}\nonumber
&&\frac{1}{(n+3)}\frac{1}{f_{3}(t)}\frac{d^{2}f_{3}}{dt^{2}} - \frac{(n+4)}{\left( n+3\right) ^{2}}\left[ \frac{1}{f_{3}(t)}\frac{df_{3}}{dt}\right] ^{2} \\&&\nonumber
+ \frac{(n-1)}{\left( n+3\right) ^{2}}\left[ \frac{1}{f_{3}(t)}\frac{df_{3}}{dt}\right] f_{1}\left( t\right) + \frac{2}{(n+3)}\frac{df_{1}}{dt} \\&&
+\frac{2\left( n+1\right) }{\left( n+3\right) ^{2}}f_{1}^{2}\left( t\right)=f_{2}(t). 
\end{eqnarray} 

On the presumption of this to be true, two point transformations are defined as 
\begin{eqnarray}\label{criterion2}
\Phi\left( T\right) &=&C\Phi\left( t\right) f_{3}^{\frac{1}{n+3}}\left( t\right)
e^{\frac{2}{n+3}\int^{t}f_{1}\left( x \right) dx },\\
T\left( \phi,t\right) &=&C^{\frac{1-n}{2}}\int^{t}f_{3}^{\frac{2}{n+3}}\left(
\xi \right) e^{\left( \frac{1-n}{n+3}\right) \int^{\xi }f_{1}\left( x
\right) dx }d\xi ,\nonumber\\
\end{eqnarray}%
which allows an exact integration of the anharmonic oscillator equation. $C$ is a constant. We work with any such cases where Eq. (\ref{integrable}), identified as a classical anharmonic oscillator equation is integrable on its own. That allows us to use the integrabilty Eq. (\ref{criterion1}) to write an additional equation for $J(z)$
\begin{eqnarray}\nonumber
&& \frac{1}{(n+3)^2 (z+1)^2 J(z)^2}\Big[-(n+4) (z+1)^2 J'(z)^2 \\&&\nonumber
+ (z+1) J(z) \Big((n+3) (z+1) J''(z) - 2 (n-3) J'(z) \Big) \\&&
+ 18 (n+1) J(z)^2 \Big] = 0.
\end{eqnarray}
This equation solvable and gives us an exact form of $J(z)$
\begin{equation}\label{capjerk}
J(z) = y (z+1)^6 \left(1-3 x (n+3) (z+1)^3\right){}^{-n-3},
\end{equation}
where $x$ and $y$ are constants of integration. Using the redefinition $2(1+j) = \phi^{(n-1)}J(z)$, we can solve for $\phi$ numerically, and from $\frac{H^2}{H_0^2} = \phi$, also derive an evolution of Hubble as a function of redshift for different values of $n$. \\
 
\begin{figure}[t!]
\begin{center}
\includegraphics[angle=0, width=0.40\textwidth]{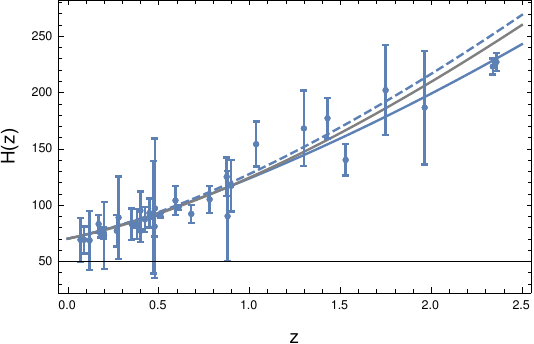}
\includegraphics[angle=0, width=0.40\textwidth]{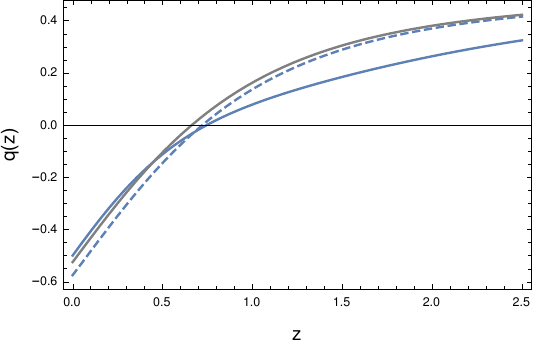}
\includegraphics[angle=0, width=0.40\textwidth]{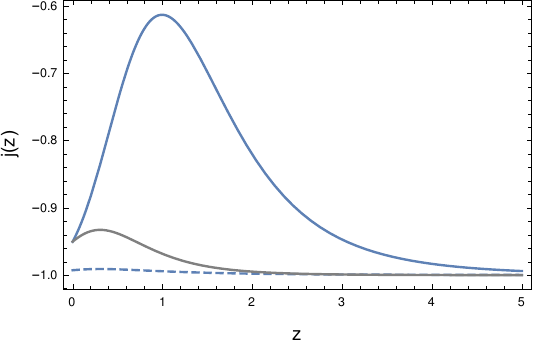}
\caption{Hubble function (Top), Deceleration parameter (Middle) and Jerk parameter (Bottom) as a function of redshift for three different parametrizations : Solid blue curve is for $n = 2$, Dashed blue curve is for $n \sim 0.001$, and Grey curve is for $n = -4$. The data points of Hubble parameter measurements (OHD) are fitted into the graph of Hubble.}
\label{hubble_q_j}
\end{center}
\end{figure}

\textit{Cosmological Implication :} 
We examine the evolution of cosmological parameters using Eq.(\ref{capjerk}). The top panel of Fig.\ref{hubble_q_j} displays the numerical solution of $H(z)$ for $n$ values : $n = 2$ (Solid blue curve), $n \sim 0.001$ (Dashed blue curve), and $n = -4$ (Grey curve). Solutions are viable for any $n$ excluding $\left\{-3,-1,0,1\right\}$. The three scenarios illustrate deviations from $\Lambda$CDM and/or observed Hubble parameter data. Graphs are fitted to observational data from direct Hubble parameter measurements (OHD) \cite{Simon_2005, Daniel_Stern_2010, Blake_2012, Moresco_2012}. We also compute the deceleration and jerk parameters as functions of redshift and plot them in the middle and bottom panels of Fig. \ref{hubble_q_j}. We also calculate the effective EoS of the cosmological system as a function of redshift, defined as $w_{eff} = \frac{p_{tot}}{\rho_{tot}}$, which can be also determined from the cosmological equations. The effective EoS is in fact, directly connected to the Hubble expansion rate through
\begin{eqnarray}
&&\frac{\rho_{tot}}{\rho_{c0}}=\frac{H^2(z)}{H^2_0},\\&&
\frac{p_{tot}}{\rho_{c0}}=-\frac{H^2(z)}{H^2_0}+\frac{2}{3}\frac{(1+z)H(z)H'(z)}{H^2_0}.
\end{eqnarray}

\begin{figure}[t!]
\begin{center}
\includegraphics[angle=0, width=0.40\textwidth]{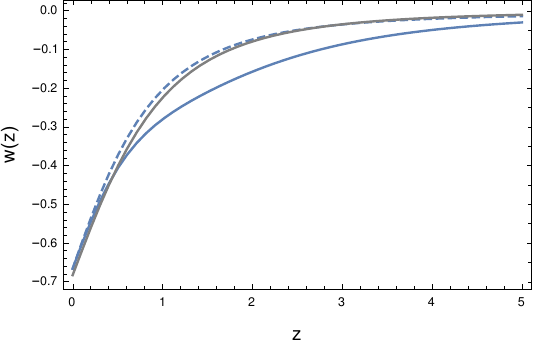}
\includegraphics[angle=0, width=0.40\textwidth]{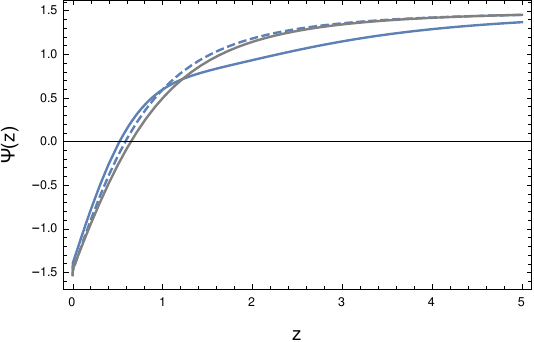}
\caption{The effective Equation of state parameter ($w_{eff}$) (Top) and Thermodynamic stability function $\Psi$ (Bottom) as a function of redshift for three different parametrizations : Solid blue curve is for $n = 2$, Dashed blue curve is for $n \sim 0.001$ and Grey curve is for $n = -4$.}
\label{w_psi}
\end{center}
\end{figure}

Here $\rho_{c0} = 3H^2_0/8\pi G$ denotes the critical density at $z = 0$. From the numerical Hubble solution, we derive $w_{eff}$ and show it in the top panel of Fig. \ref{w_psi}. The Hubble and deceleration plots may indicate that the late-time evolution resembles $\Lambda$CDM; however, the jerk evolution reveals a noticeable deviation from standard $\Lambda$CDM (for which $j = -1$).     

It is important that any reconstructed cosmological model must evolve into a thermodynamic stability. Much like a spherical black hole in thermodynamic equilibrium \cite{PhysRevD.15.2738, PhysRevLett.75.1260}, the spherical, expanding cosmological universe can be imagined as a system enclosed by the Hubble horizon \cite{Bak_2000}, defined as $r_h = (H^2+k/a^2)^{-1/2}$. By defining the total entropy of the system $S$ as a sum of boundary entropy and fluid entropy, i.e., $S = S_f + S_h$ \cite{PhysRevD.15.2738}, we analyze the thermodynamics stability through a Hessian matrix analysis, leading to the conditions $\frac{dS}{dN} \geq 0$ and $\frac{d^2S}{dN^2} < 0$ where $N = \ln{a}$. The horizon entropy is proportional to the horizon area. Under the assumption that the constituent fluid elements of the universe have a uniform temperature $T$ \cite{PhysRevLett.75.1260}, the effective fluid entropy can be defined as $S_f = S_{cdm} + S_{de}$. Using the first law of thermodynamics we derive that the first and second order change in entropy can be written as a function of Hubble \cite{10.1093/mnras/stab1910, 10.1093/mnras/stac979} as
\begin{align}\notag
S_{,N} &= \frac{16\pi^2}{H^4}(H_{,N})^2, \\ \label{Sdn}
S_{,NN} &= 2S_{,N}\left(\frac{H_{,NN}}{H_{,N}}-\frac{2H_{,N}}{H}\right) \equiv \Psi.
\end{align}

We note that a thermodynamic equilibrium requires $\Psi$ to be negative and this can easily be checked for a reconstructed model. We show $\Psi$ as a function of redshift in the bottom panel of Fig. \ref{w_psi} for three chosen values of $n$ and infer that $\Psi$ always evolves into negative values at late times. Therefore, the universe has a proclivity to move towards thermodynamic equilibrium with its expansion. 

The clue from Fig. \ref{hubble_q_j} is that the numerical solution matches with OHD data most accurately when $n$ is very close to zero (from either side). One can further confirm this from the jerk, whose evolution becomes arbitrarily close to $\Lambda$CDM for $n \to 0$. However, a direct choice $n = 0$ is restricted from the analysis and therefore, we take up the $n \to 0$ limiting case. In this limit, we solve Eqs. (\ref{jerk1}) and (\ref{capjerk}) simultaneously  
\begin{equation}
\phi''(z) -\frac{2 \phi'(z)}{(z+1)}+\frac{2 \phi(z)}{(z+1)^2} \left\lbrace \frac{y(z+1)^6}{2\phi(z)\left(1- 9x(z+1)^3 \right)^3} \right\rbrace = 0,
\end{equation}

and derive the dimensionless Hubble function $\frac{H}{H_{0}} = E(z)$ as
\begin{equation}\label{E}
E(z)^{2} = \left[\frac{c_1}{3} (1+z)^3 - \frac{2c_1}{54x} + \frac{y}{1458 x^2 \left\lbrace -1+9x(1+z)^3 \right\rbrace}\right].
\end{equation} 

We reframe the above equation in the following form,
\begin{equation}\label{eq:E}
E(z)^{2} = \left[\Omega_{m0} (1+z)^3 - \frac{\Omega_{m0}}{\beta} + \frac{\alpha}{\beta^2 \left(\beta (z+1)^3 - 1 \right\rbrace}\right].
\end{equation} 

$\Omega_{m0}=\frac{c_1}{3}$, $\alpha = \frac{y}{18}$ and $\beta = 9x$. Using the requirement that at $z = 0$, $H = H_0$ we connect the parameters $\Omega_{m0}, \alpha$ and $\beta$ through a constraint equation
\begin{equation}\label{eq:constrain}
\alpha = (\beta -1) \beta ^2 \left(\frac{\Omega_{mo} }{\beta }-\Omega_{m0} +1\right),
\end{equation}

and re-write the Hubble function $H(z) = H_0 E(z)$ as 

\begin{eqnarray}
&& H(z)^2= H_{0}^{2}\left[\Omega_{m0}(1+z)^3 + \Omega_{DE}\right], \\&&
\Omega_{DE} = - \frac{\Omega_{m0}}{\beta} + \frac{\alpha }{\beta ^2 \left(\beta  (z+1)^3 - 1 \right\rbrace}.
\end{eqnarray}

The corresponding dark energy EoS is derived as
\begin{equation}
    \begin{array}{l}
w_{\phi}(z)=\frac{-1-\frac{2 \dot{H}}{3 H^{2}}}{\Omega_{\phi}}=\frac{\frac{2}{3}(1+z) E \frac{d E}{d z}-E^{2}}{E^{2}-\Omega_{m 0}(1+z)^{3}},
\end{array}
\end{equation}

The dark energy EoS for the normalized Hubble parameter $E(z)=\frac{H(z)}{H_0}$ is derived as
\begin{equation} \label{eq:eos}
w = -1 + \frac{\frac{ \alpha \beta a^{-3} }{ \beta^2 \left( \beta a^{-3} - 1 \right)^2 })}{ \left( \frac{\Omega_{m,0}}{\beta} - \frac{\alpha}{\beta^2 \left( \beta a^{-3} - 1 \right)} \right)}.  
\end{equation}

It is interesting to note that after using the constraint Eq.(\ref{eq:constrain}), the dark energy EoS involves only one parameter $\beta$. \\

\begin{figure}
    \centering
    \includegraphics[width=\linewidth]{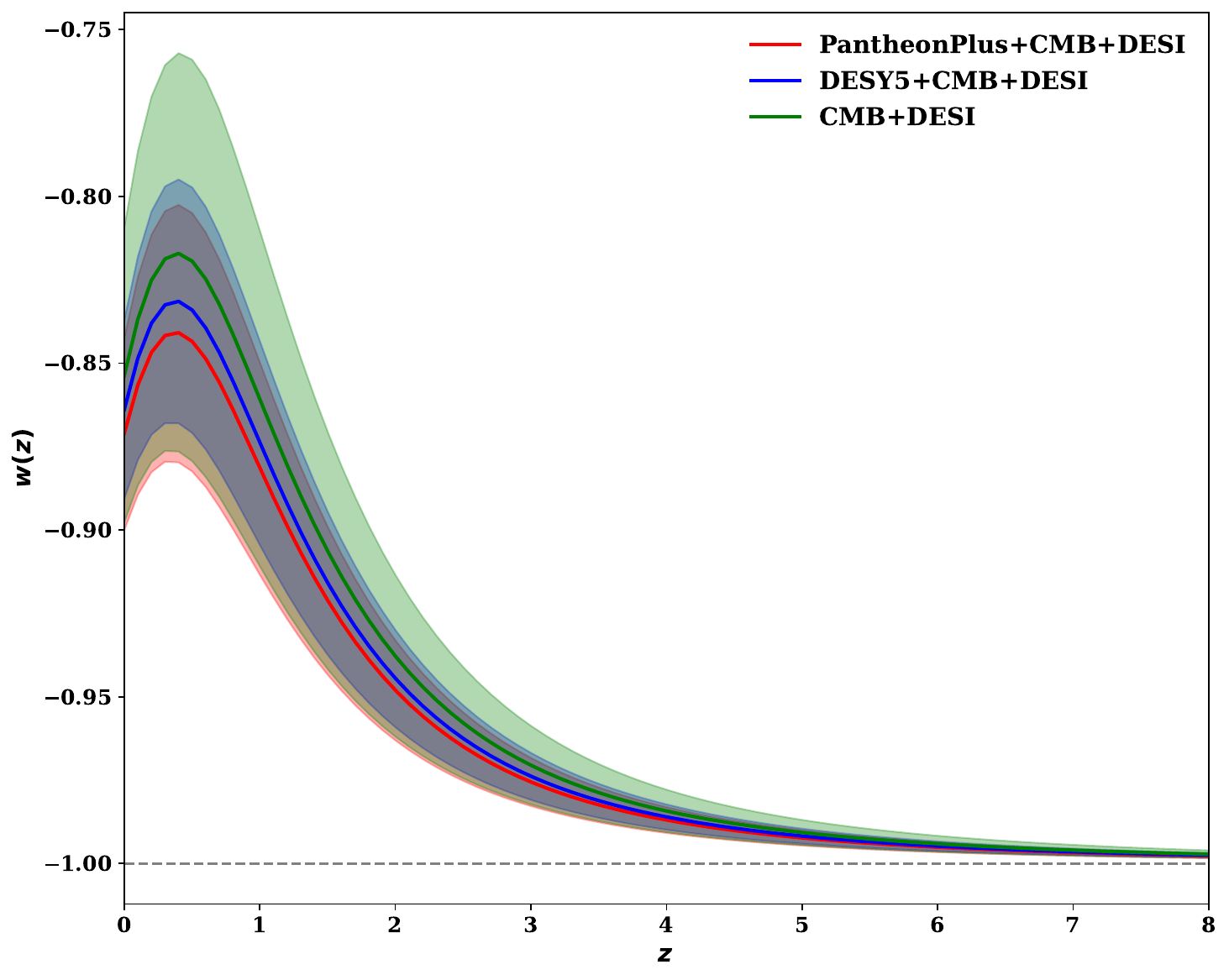}
    \caption{The plot shows the evolution of the dark energy equation of state. The solid red, blue, and green lines represent datasets \setone, \settwo, and \setthree, respectively. The corresponding dark and light shaded areas indicate the $1\sigma$ and $2\sigma$ confidence intervals.}
    \label{fig:eos}
\end{figure}

\textit{Observational Data:}
We compare the dark energy EoS given in Eq.(\ref{fig:eos}) against the available cosmological data using a Markov Chain Monte Carlo (MCMC) analysis and the plotting tool GetDist to generate posterior distributions. We use the Pantheon Plus compilation of SN-Ia data~\cite{Scolnic:2021amr, Riess_2022, Brout:2022vxf, Brout:2022vxf, Riess:2021jrx} alongside the DES Year 5 dataset~\cite{DES:2024jxu}. These are characterized by distinct photometric systems and selection criteria and both of them provide distance moduli $\mu$ for a range of redshifts. We utilize the 2025 BAO observational data from the Dark Energy Spectroscopic Instrument (DESI-DR2) as reported in~\cite{DESI:2025zgx, DESI:2024mwx} and the Planck18 compressed likelihood ~\cite{Chen:2018dbv} (henceforth referred as CMB). The compressed CMB prior offers a convenient substitute for a complete global fit of the entire Planck dataset. We categorically use the following three combinations of datasets to constrain the kinematic model : \setone, \settwo and \setthree. \\

\begin{figure}[t!]
    \centering
    \includegraphics[width=\linewidth]{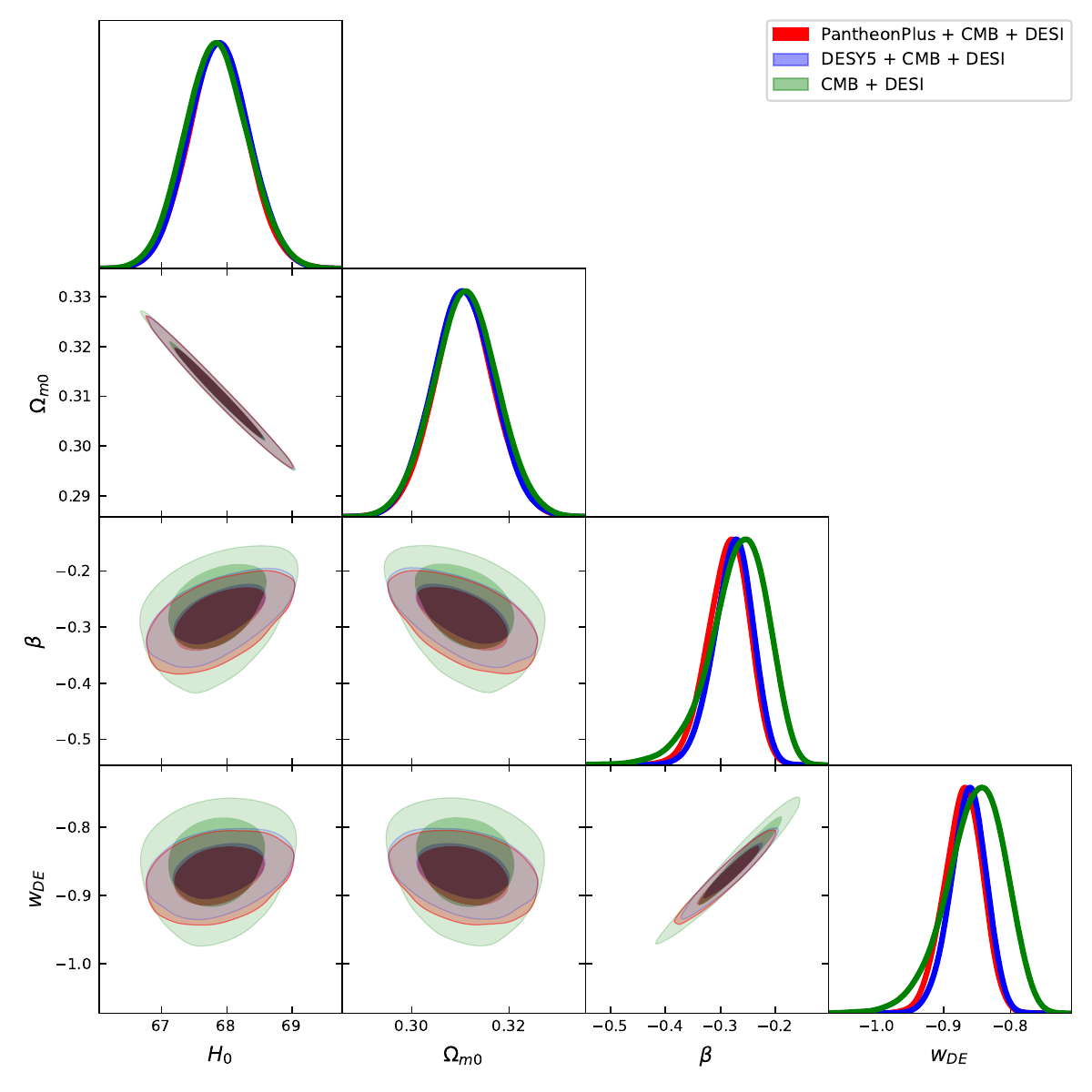}
    \caption{2D and 1D marginalized posterior distributions (triangle plot) for the cosmological parameters $H_0$, $w_b h^2$, $\Omega_{m0}$, and $r_d$, along with the model parameter $\beta$ and the present value of the dark energy equation of state $w_{DE}$. The red contours correspond to the \textit{PantheonPlus + CMB + DESI DR2} dataset, the blue to \textit{DESY5 + CMB + DESI DR2}, and the green to \textit{CMB + DESI DR2}.}
    \label{fig:triangle}
\end{figure}

\textit{Results:}
We impose flat prior on the following parameters, $\Omega_{m0}:[0.2,0.4], H_0:[60,80], r_d:[80,200],\beta:[-1,-0.1]$. It is necessary to choose $\beta < 0$ from a numerical standpoint, as any $\beta > 0$ leads to numerical instabilities while integrating the system. We report the mean values of the cosmological parameters obtained from our analysis in Table:\ref{tab:bestfit} along with the corresponding $1\sigma$ constraints. In Fig.\ref{fig:triangle} we show the 2D and 1D triangular plot of the cosmological parameters $H_0, \Omega_{m0}$, the model parameter $\beta$, and the current value of the dark energy EoS, $w_{DE}$. The red plot represents the posteriors for \textit{\setone} data, the blue corresponds to the \textit{DESY5 + CMB + DESI DR2} while the green displays results for the \textit{CMB + DESI DR2} combination. It can be seen from Fig. \ref{fig:triangle} that the posteriors from all these data sets resemble excellent consistency with each other. \\

\begin{table*}[ht]
\centering
\begin{tabular}{|l|cc|cc|cc|}
\hline
Parameters 
& \multicolumn{2}{c|}{Pantheon Plus+CMB+DESI DR2}
& \multicolumn{2}{c|}{DES Y5+CMB+DESI DR2}
& \multicolumn{2}{c|}{CMB+DESI DR2} \\
\hline
& $\Lambda$CDM & $w$CDM 
& $\Lambda$CDM & $w$CDM 
& $\Lambda$CDM & $w$CDM  \\
\hline
$\Omega_{m0}$ 
    & $0.3178\pm 0.0058$ 
    & $0.3107\pm 0.0062$
    & $0.3195\pm 0.0058$
    & $0.3106\pm 0.0063$ 
    & $0.3129\pm 0.0060$ 
    & $0.3111\pm 0.0065$  \\
$H_0$ 
    & $67.36\pm 0.42$ 
    & $67.87\pm 0.46$
    & $67.24\pm 0.42$
    & $67.88\pm 0.46$ 
    & $67.71\pm 0.44$ 
    & $67.84\pm 0.48$  \\
$r_d$ 
    & $147.80\pm 0.43$ 
    & $143.2\pm 1.3$ 
    & $147.87\pm 0.43$
    & $142.9\pm 1.3$
    & $147.59\pm 0.44$ 
    & $142.5\pm 1.9$ \\
$\Omega_b h^2$ 
    & $0.02237\pm 0.00013$ 
    & $0.02245\pm 0.00014$
    & $0.02234\pm 0.00013$
    & $0.02245\pm 0.00013$  
    & $0.02242\pm 0.00013$ 
    & $0.02244\pm 0.00014$  \\
$w_{\phi}$ 
    & $-1$ & $-0.871^{+0.031}_{-0.026}$ 
    & $-1$ & $-0.864^{+0.028}_{-0.025}$
    & $-1$ & $-0.854^{+0.050}_{-0.037}$ \\
$\beta$ 
    & $-$ & $-0.286^{+0.041}_{-0.034}$ 
    & $-$ & $-0.278^{+0.039}_{-0.032}$
    & $-$ & $-0.268^{+0.061}_{-0.041}$ \\
$\Delta \chi^{2}$ 
    & $0$ & $-16.826$
    & $0$ & $-19.8955$
    & $0$ & $-9.096$ \\
$\Delta AIC$
    & $0$ & $-14.826$
    & $0$ & $-17.8955$
    & $0$ & $-7.096$ \\
\hline
\end{tabular}
\caption{Mean values of different cosmological parameters together with $68\%$ constraints. The result for the $\Lambda$CDM model is reported for comparison.}
\label{tab:bestfit}
\end{table*}

\begin{figure}
    \centering
    \includegraphics[width=\linewidth]{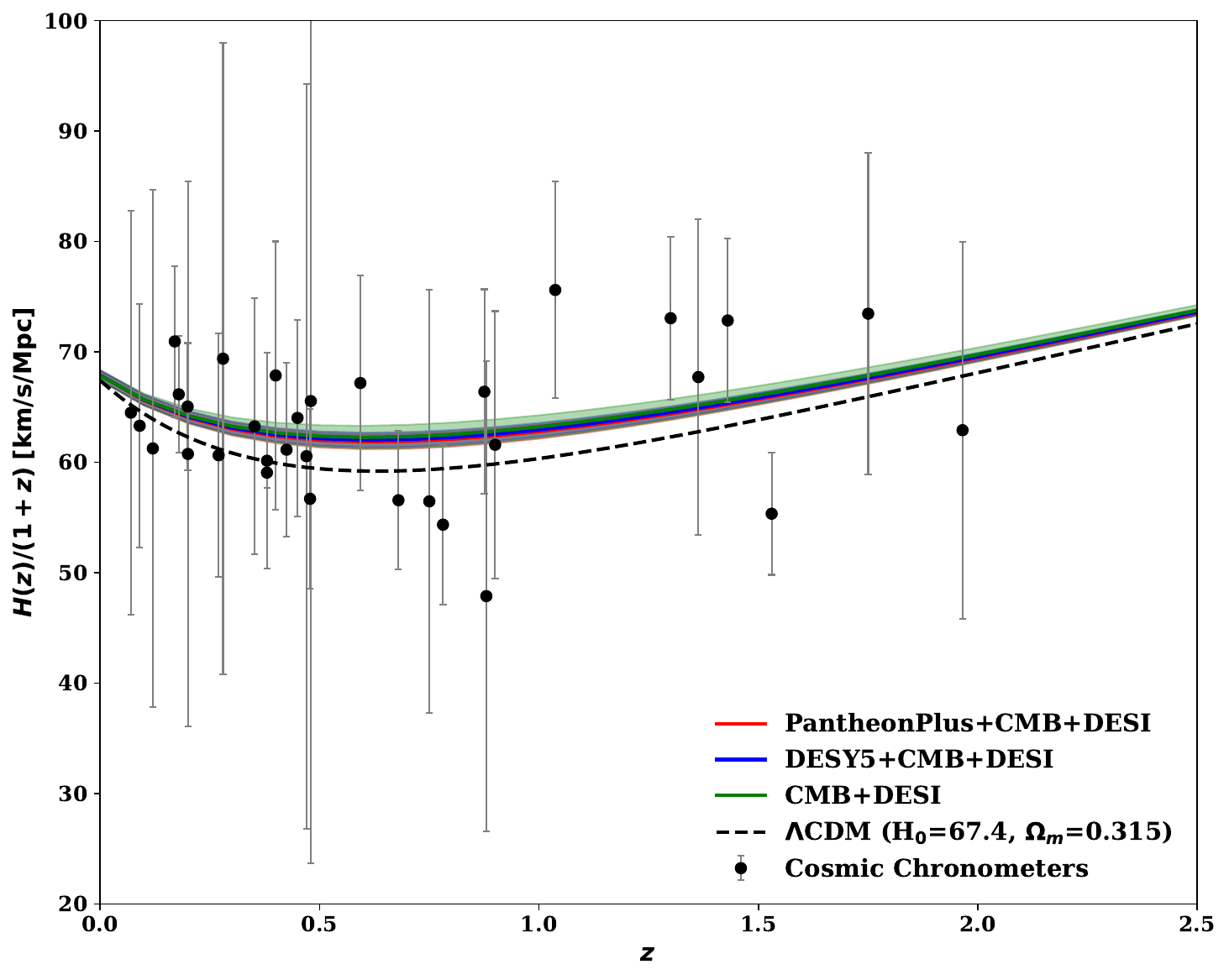}
    \caption{Evolution of $H(z)/(1+z)$ against the redshift $z$. Data is taken from \cite{Favale:2023lnp}. The color legend corresponds to the same dataset combinations as in the previous plot.}
    \label{fig:Hubble_posterior}
\end{figure}

\begin{figure}
    \centering
    \includegraphics[width=\linewidth]{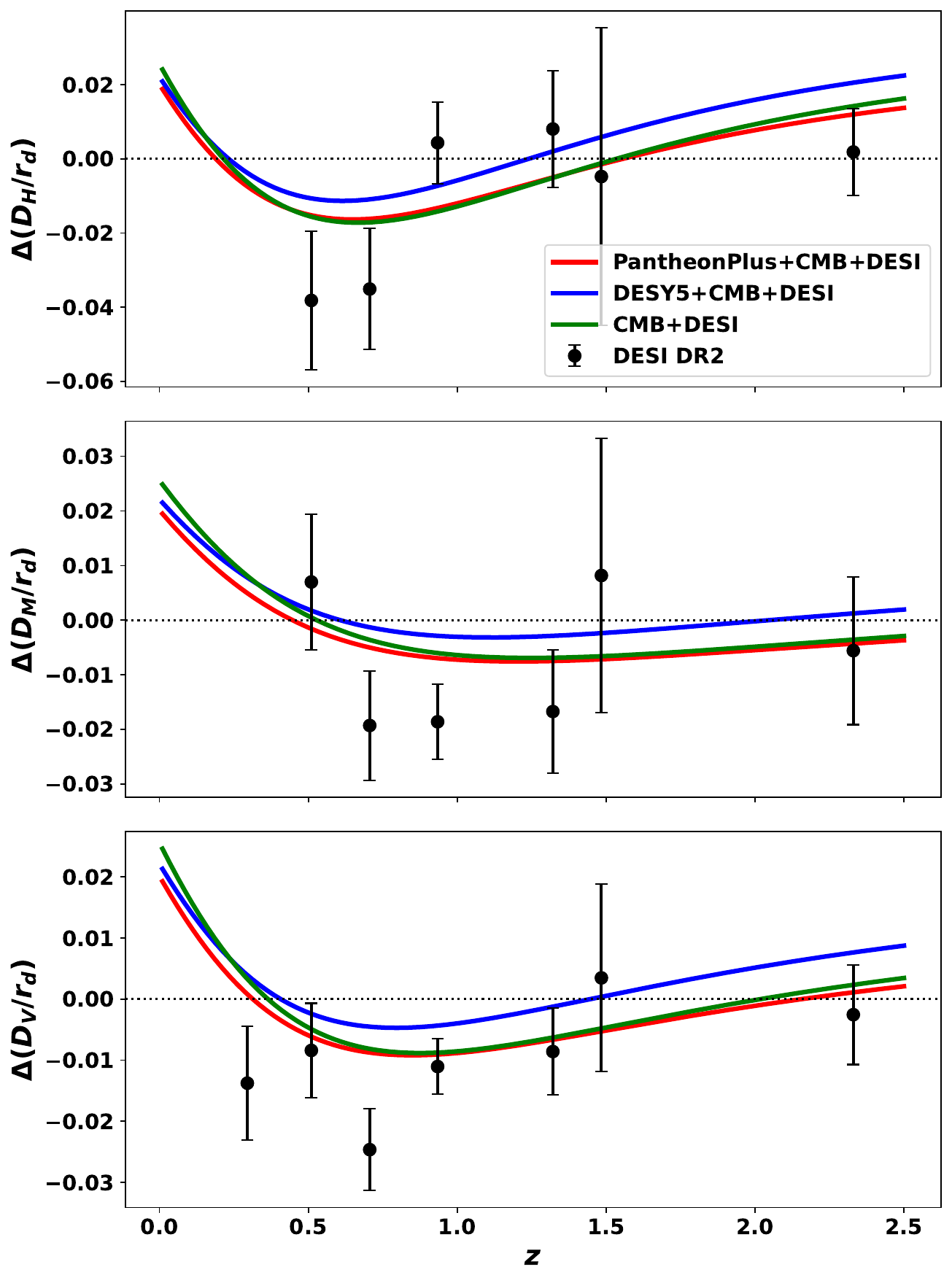}
    \caption{Differential plots of $D_H/r_d$, $D_M/r_d$, and $D_V/r_d$ relative to $\Lambda$CDM. The color legend corresponds to the same dataset combinations as in the previous plot.}
    \label{fig:placeholder}
\end{figure}

We evaluate the statistical performance of our model in comparison to the $\Lambda$CDM, using the minimum chi-squared value ($\chi^2_{\text{min}}$), the chi-squared difference ($\Delta \chi^2$) and the Akaike Information Criterion difference ($\Delta \text{AIC}$). We show the relative difference in minimum chi-square values between our model and the $\Lambda$CDM model in Table-\ref{tab:bestfit}. Across each dataset, our model demonstrates a better fit to the observational data compared to the $\Lambda$CDM. From Table \ref{tab:bestfit}, one can see that for \setone, the proposed model has the lowest value of $\chi^2_{\text{min}}$. We evaluate the Akaike Information Criterion (AIC), defined as $\Delta{\rm AIC} = \chi^{2}_{\min,\mathcal{M}} - \chi^{2}_{\min,\Lambda{\rm CDM}} + 2(N_{\mathcal{M}}-N_{\Lambda{\rm CDM}})$. A negative $\Delta{\rm AIC}$ favors $\mathcal{M}$, with larger magnitude indicating stronger support. For the data sets \setone, \settwo, and \setthree, we obtain $\Delta{\rm AIC} = -14.862$, $-17.896$, and $-7.096$, respectively, providing clear preference for our model over $\Lambda$CDM. In particular, the first two datasets, which include supernova observations, provide strong evidence for our model, while the third offers moderate support, consistent with DESI results. Hence, the dynamical nature of the dark energy is once again supported over the cosmological constant.  \\


We plot the dark energy EoS in Fig.~\ref{fig:eos} directly from the MCMC chains generated from our analysis. The solid red line corresponds to the dataset \setone, the blue line to \settwo and the green line to \setthree. The dark and light shaded regions in each corresponding colour represent the $1\sigma$ and $2\sigma$ confidence level contours. We see that in the proposed model, there is no phantom barrier crossing that occurred in the recent past, although our analysis shows that the current value of the $w_{DE}$ is in the quintessence region. Therefore, our results oppose the DESI~\cite{DESI:2025zgx, DESI:2024mwx} analysis that employs the CPL parametrization where a phantom barrier crossing has been reported. In Fig.\ref{fig:Hubble_posterior} we show the Hubble parameter evolution for our model, which is consistent with the observational data. We also present differential plots of $D_H/r_d$, $D_M/r_d$, and $D_V/r_d$ relative to $\Lambda$CDM, together with their evolution in our model. These comparisons indicate that our model provides a better fit to the DESI-DR2 measurements than $\Lambda$CDM. The novelty of our approach lies in the fact that it avoids arbitrary choices of $w_{DE}$ parametrization. Instead, we adopt an analytic solution of the differential equation for the jerk parameter, leading to a description governed by a single parameter.
 
\begin{figure}
    \centering
    \includegraphics[width=\linewidth]{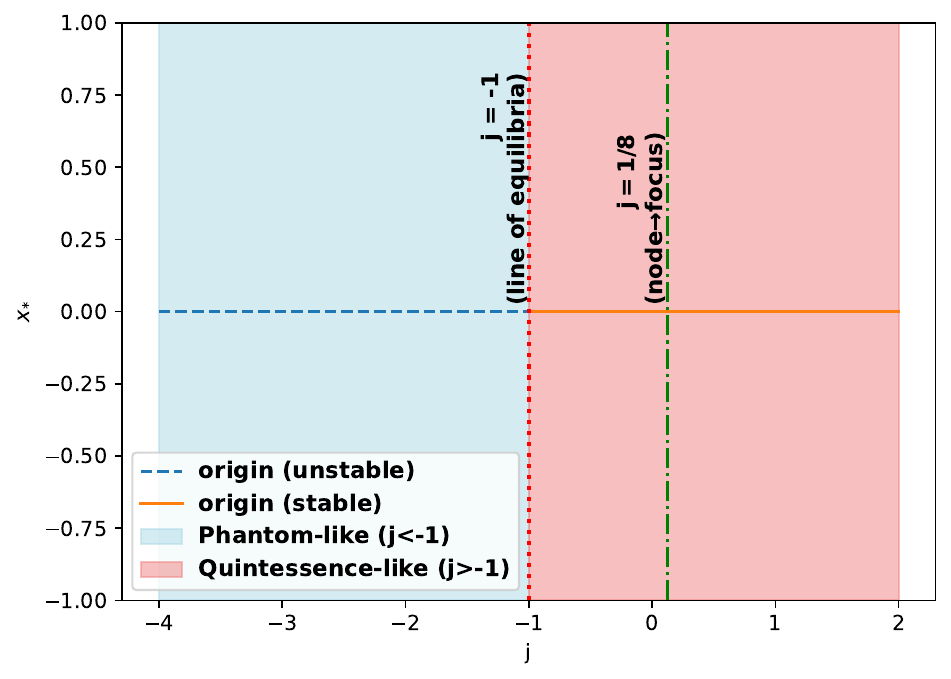}
    \caption{Bifurcation diagram for the dynamical system given in Eq.\ref{eq:ds}.}
    \label{fig:bifurcation}
\end{figure}

Before concluding the letter, we give a dynamical systems perspective of our arguments, mainly related to the absence of a phantom barrier crossing in the dark energy EoS. For a toy dark energy model with a constant EoS parameter $w_0$, it is easy to express the jerk parameter as $
j = -1 - \frac{3}{h^2}(1-\Omega_0)\, w_0(1+w_0)(1+z)^{3(1+w_0)}$ and verify that for quintessence-like models ($-1 < w_0 < 0$) the $j > -1$, while for phantom models ($w_0 < -1$) $j < -1$. The $\Lambda$CDM ($w_0=-1$) corresponds exactly to $j=-1$. Therefore any transition across the phantom barrier necessarily requires the jerk parameter to cross a critical value $j=-1$ which is simply a bifurcation of the underlying dynamical system. To demonstrate this we rewrite the cosmographic identity in Eq.(\ref{jerk2}) as a set of two first-order differential equations.

\begin{equation}\label{eq:ds}
\frac{dy}{dN} = x ~,~ \frac{dx}{dN} = -3x - 2(j+1)y,
\end{equation}

where $x=\dot{\phi}$ and note that the only fixed point is $(x,y)=(0,0)$ when $j \neq -1$. At $j=-1$, however, the entire $y$-axis becomes an equilibrium manifold. The fixed point is stable for $j>-1$ and unstable for $j<-1$, with a bifurcation occurring precisely at $j=-1$. Since $j=-1$ also corresponds to the $\Lambda$CDM case ($w_0=-1$), the bifurcation implies that any smooth crossing of the phantom divide must pass through this point. However, because $j=-1$ behaves as an attractor, no trajectory can cross it. Therefore, we note that from the dynamical systems perspective, phantom divide crossing is dynamically prohibited within a cosmographic framework. This is illustrated in the Fig.\ref{fig:bifurcation}. Interestingly, a bifurcation also arises in the spatial curvature at the phantom barrier. In a non-flat universe, the relation between the jerk and curvature is $\Omega_k = \Omega_{m0}(a^{-4}-1) - (j+1)/a$ \cite{Amirhashchi:2018vmy} with flatness achieved when $\Omega_{m0}(a^{-4}-1) = (j+1)/a$. For $j>-1$, $\Omega_k>0$, while $j<-1$ yields $\Omega_k<0$, signaling a curvature bifurcation. Notably, this result follows directly from the cosmographic definitions, independent of the specific analytic solution obtained in this analysis.
\\

\textit{Conclusion:}
The letter includes a few important arguments in the context of the recent findings from DESI DR2. Usually it is understood that a CPL parameterization prescribes a preference for dynamical dark energy and allows a transition of the universe from the phantom regime to the quintessence domain in recent past. However, there remains questions regarding whether this phantom barrier crossing is a genuine physical phenomenon or merely an artefact of the data or the specific dark energy parameterization employed. We propose a dynamical dark energy model derived analytically from the differential equation associated with the cosmographic jerk parameter. Initially, this differential equation is formulated using the definitions of cosmographic parameters and is subsequently mapped as an anharmonic oscillator equation. We employ a point transformation which allows us to close the differential equation and find an exact solution. We show the numerical solutions of the system for different cosmological parameters and examine the background dynamics. Furthermore, for the ground state of the anharmonic oscillator, we derive a solution leading to a one-parameter parametrization of the dark energy EoS.  We perform an MCMC analysis using currently available data to constrain the model parameters for three different combinations of the data sets, \setone, \settwo, and \setthree. \\

Our findings suggest that although the behaviour of all cosmological parameters align with theoretical expectations and observations favour dynamical dark energy, there is no evidence of a phantom barrier crossing for the dark energy EoS in the past. In that sense, our results contradict the DESI results. The formulation enables us to probe the dynamics of dark energy directly from fundamental kinematical quantities without imposing specific phenomenological constraints. Dynamical system analysis shows that the phantom barrier is a bifurcation point, prohibiting solutions from crossing it. These results underscore the importance of model-independent reconstructions in understanding the true dynamics of dark energy in current and future high-precision cosmological surveys.

\section*{Acknowledgement}

The authors acknowledge Roy Maartens for his comments on the article. SC acknowledges the IUCAA for providing facility and support under the visiting associateship program. Acknowledgement is given to the Vellore Institute of Technology for the financial support through its Seed Grant (No. SG20230027), 2023.

\bibliographystyle{unsrt}
\bibliography{jerk}

\end{document}